# Determination of the STM tip-graphene repulsive forces by comparative STM and AFM measurements on suspended graphene†

András Pálinkás,[a,c] György Molnár,[a] Chanyong Hwang,[b,c] László Péter Biró,[a,c] and Zoltán Osváth[a,c,*]

[a]Institute of Technical Physics and Materials Science (MFA), Centre for Energy Research, HAS, 1525 Budapest, P.O. Box 49, Hungary
[b]Center for Nano-metrology, Korea Research Institute of Standards and Science, Yuseong, Daejeon 305-340, South Korea
[c]Korea-Hungary Joint Laboratory for Nanosciences (KHJLN), 1525 Budapest, P.O. Box 49, Hungary
†Electronic supplementary information (ESI) available. See DOI: 10.1039/c6ra19660h
*E-mail: zoltan.osvath@energia.mta.hu

## Abstract

Graphene grown by chemical vapour deposition is transferred on top of flat gold nanoislands and characterized by scanning tunnelling microscopy (STM) and atomic force microscopy (AFM). Graphene bubbles are formed with lateral dimensions determined by the size and shape of nanoislands. These graphene bubbles can be squeezed during STM imaging using bias voltages of less than 250 mV and tunnelling currents of 1 nA. Similarly, the graphene suspended over gold nanovoids is deflected 4 – 5 nm by the STM tip when imaging at low bias voltages (U = 30 mV). Nanoindentation measurements performed by AFM show that the squeezing of graphene bubbles occurs at repulsive forces of 20 – 35 nN, and such forces can result in deflections of several nanometres in suspended graphene parts, respectively. Comparing the AFM and STM results, this study reveals that mechanical forces of the order of $10^{-8}$ N occur between the STM tip and graphene under ambient imaging conditions and typical tunnelling parameters.

## Introduction

The role of mechanical forces in the interpretation of scanning tunnelling microscopy (STM) images was addressed soon after the invention of this imaging technique.[1] Long-range van der Waals forces dominate the STM tip-sample interaction at separations larger than 1 nm. As the tip-sample distance decreases, strong attractive adhesion forces appear for tip-surface separations larger than the equilibrium distance. If the tip approaches more closely the sample, the quantum-mechanical forces become repulsive, as a consequence of the Pauli exclusion principle.[2] Early atomic resolution experiments performed on cleaved graphite surfaces in constant current mode demonstrated giant corrugation amplitudes of several Ångstroms.[3,4] It was proposed that elastic deformations induced by the interatomic forces between STM tip and surface are responsible for the increased electronic corrugation.[4] Direct force measurements during STM imaging of graphite in air revealed high repulsive tip-sample forces in the range of $10^{-7} – 10^{-6}$ N,[5,6] which were attributed to the presence of a surface contamination layer.[7] Later, the attractive adhesion force regime was also observed by using a flexible cantilever beam as sample stage during normal metal-vacuum-metal tunnelling operation of the STM.[8,9] More recently it was shown that both the van der Waals force and an electrostatic force from the STM tip can induce substantial mechanical deformation in suspended graphene membranes,[10–12] which indicates possible applications for generating controlled strain and pseudomagnetic fields in graphene.[13,14] Similarly, mechanical distortions of suspended graphene can be provoked also in the repulsive force regime of the STM tip-graphene interaction.[15–17] Here we show that graphene bubbles formed on flat gold nanoislands can be squeezed by STM imaging in the repulsive force regime, and also that graphene suspended over gold nanovoids can be deflected by the STM tip. Comparing the STM-induced deflections to the deflections induced by AFM nanoindentation experiments we were able to quantify the repulsive forces of the STM tip-graphene interaction.

## Experimental

Gold nanoislands were prepared by evaporating 5-8 nm of gold (99.99% purity) onto highly oriented pyrolytic graphite (HOPG) substrates at a rate of 0.1 nm/s. After evaporation the samples were annealed at 400 °C in argon atmosphere for 1 hour, which resulted in the formation of gold nanoislands through the surface diffusion and coalescence of gold clusters. The nanoislands have typically heights of 15-20 nm and lateral dimensions of several hundreds of nanometres, as observed by AFM measurements.

Large-area graphene was grown by chemical vapour deposition (CVD) on a mechanically and electro-polished copper foil, as described elsewhere.[18] The graphene sample was transferred onto the gold nanoislands using thermal release tape, and copper etchant.[18] After the etching procedure, the tape holding the graphene was rinsed in distilled water, then dried and pressed onto the HOPG surface decorated by the gold nanoislands. The sample was placed on a hot plate and heated to 100 °C, which is above the release temperature (90 °C) of the tape. The tape was easily removed at this temperature, and graphene flakes were successfully transferred on top of the gold nanoislands (and on the HOPG). A second annealing was performed at 400 °C in argon atmosphere for 30 min, to improve the adhesion of graphene to the nanoislands.

The graphene-covered gold nanoislands were investigated by AFM, STM and STS, using a Bruker MultiMode 8 and a DI Nanoscope E operating under ambient conditions. We used mechanically-cut Pt/Ir (90/10%) tips in the STM experiments. The AFM measurements were performed in both Tapping and PeakForce® mode.[19] The PeakForce mode is a relatively new scanning mode, where the probe and sample are intermittently brought together (similar to Tapping Mode) to contact the surface for a short period, which eliminates lateral forces. A complete force-distance curve is performed in every measuring point of the scanned area, while the z-piezo data of the cantilever is recorded at the maximum repulsive force (Peak Force) between the sample and the cantilever.[19] The maximum force can be changed in order to acquire images at different sample-cantilever forces. For the indentation experiments on suspended graphene, we used an AFM cantilever with maximal tip radius of R = 12 nm (RTESPA model from Bruker) and spring constant k = 20.5 N/m, as determined in situ by the thermal tune method.[20]

## Results and discussion

### *Graphene nanobubbles*

The first sample that we investigated was prepared by evaporating 5 nm of gold on HOPG, and subsequent annealing at 400 °C. Figure 1a shows an STM image of the resulting gold nanoislands, after graphene transfer (see also Figure S1 in the supplementary information). The nanoislands are flat, irregularly shaped, well separated from each other, typically of 20 nm in height, and lateral dimensions of several hundreds of nanometres, or sometimes even going beyond 1 μm.

Most of the area shown in Figure 1a is covered with graphene. Note that quasi-one-dimensional nanostructures can be observed on the surface (arrows in Fig. 1a), which often connect neighbouring nanoislands. These nanostructures are partly gold nanowires, and partly graphene wrinkles (see Figure S2, supplementary information). More interestingly, huge graphene bubbles form on the top of gold nanoislands (marked with dots in Fig. 1a), implying large regions of suspended graphene. These graphene bubbles are very sensitive to the mechanical forces arising from the proximity of the STM tip. To illustrate this, we turn the attention to the graphene bubble in Figure 2. Here, subsequent STM images of the same bubble are shown, which were acquired at constant tunnelling current (I = 1 nA). The nanobubble can be characterized with a length (*l*), width (*w*), and height (*h*) of 610 nm, 405 nm, and 44 nm, respectively. In Figure 2a, a relatively large bias voltage of U = 1 V was used. We think that at this bias voltage the tip-sample repulsive forces are minimized and the attractive force regime dominates the STM tip-graphene mechanical interaction. The STM image of the graphene bubble is noisy, which is typical when suspended graphene is measured by STM.[21] The height profile corresponding to the white line in Fig. 2a is shown in Fig. 2d (black solid line). Similar height profile along the same line is obtained when the bubble is imaged at U = 500 mV (Fig. 2d, black dashed line).

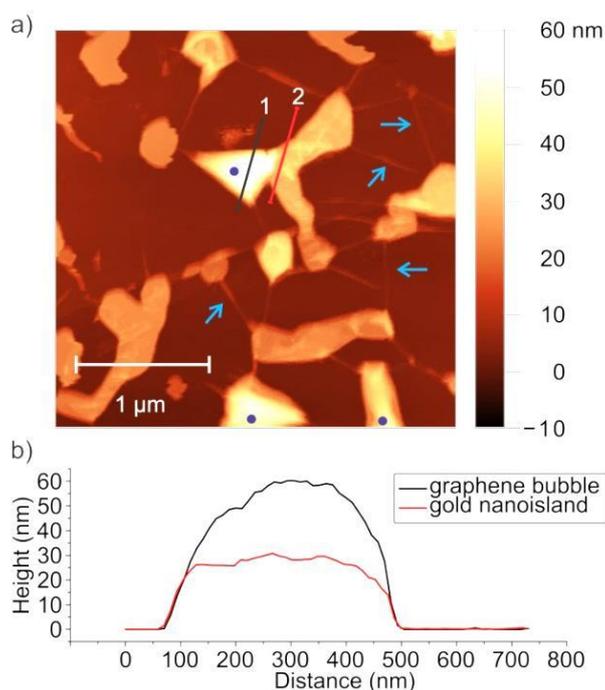

Figure 1. (a) STM image of graphene covered gold nanoislands. Graphene wrinkles connecting neighbouring nanoislands are marked with arrows. Graphene bubbles form on gold nanoislands (marked by dots). (b) Height profiles corresponding to the line sections 1 and 2 in a).

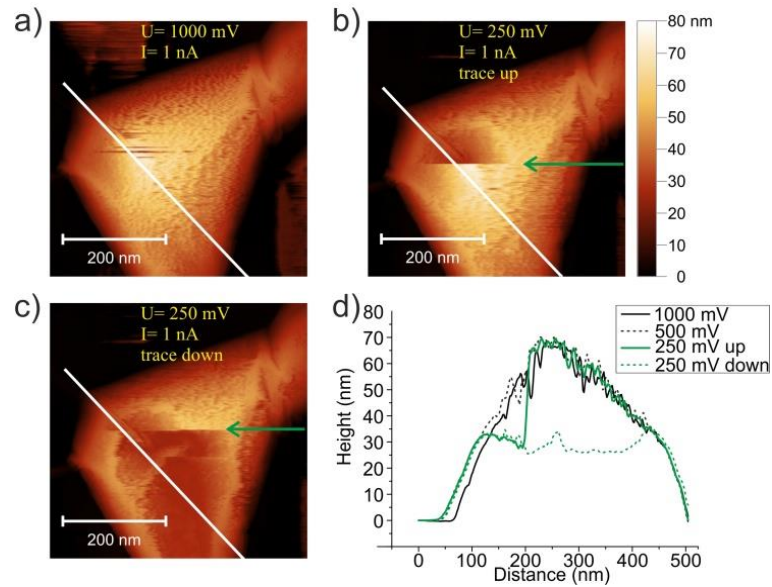

Figure 2. STM images of a graphene nanobubble measured at bias voltages of (a) U = 1000 mV, (b) U = 250 mV, slow scan axis from bottom to top (trace up), and (c) U = 250 mV, slow scan axis from top to bottom (trace down). Collapse of the nanobubble is observed at the scan lines marked with green arrows in b) and c). Tunneling current: I = 1 nA. (d) Height profiles taken along the same line section shown in a)-c) (white line).

However, when we decreased the bias voltage to U = 250 mV the bubble suddenly collapsed when imaging the centre of the bubble (Fig. 2b, at the line marked with arrow). This is shown also on the height profile in Fig. 2d (green solid line). A subsequent STM image performed with the same bias voltage of 250 mV shows that graphene was further pushed downwards (Fig. 2c, at the line marked with arrow), until it reached the surface of the underlying gold nanoisland. Thus, the corresponding height profile (Fig. 2d, green dashed line) gives the height of the gold nanoisland on which the graphene bubble initially formed. Increasing the bias voltage again to 1000 mV, the graphene bubble does not recover its initial shape (see Figure S3, supplementary information). In this case, the graphene undergoes an abrupt switching from convex to concave geometry. Such switching was observed recently when electrostatic pressure was applied to a graphene membrane via an underlying gate electrode.[22] The switching occurs at a critical pressure, when energetically becomes more favourable to form a concave shape.[22] Similar STM tip induced deformations were observed on the graphene bubble shown in Figure 3 as well (*l*, *w* and *h* of 625 nm, 370 nm, and 27 nm).

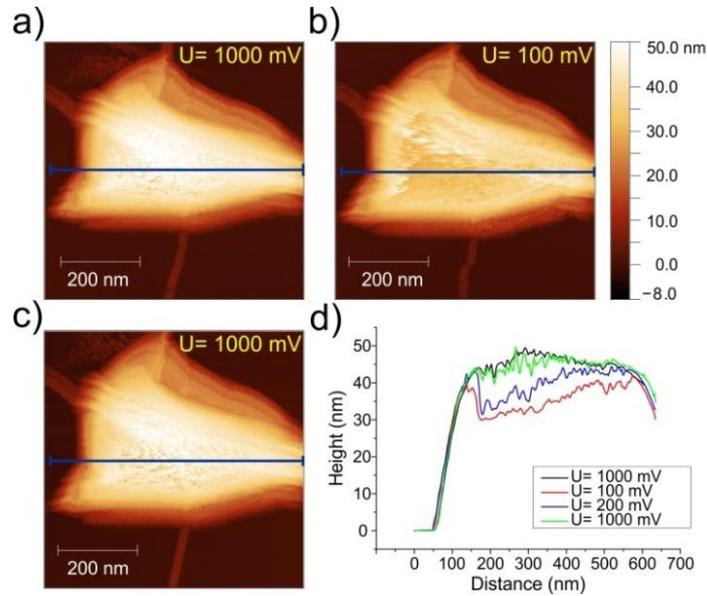

Figure 3. Series of STM images of a graphene nanobubble measured at bias voltages of (a) U = 1000 mV, (b) U = 100 mV, and (c) U = 1000 mV. Tunneling current: I = 1 nA. (d) Height profiles taken at different bias voltages along the same line section shown with horizontal line in a)-c).

First, the graphene bubble is imaged at U = 1000 mV (Figure 3a), like in the previous case. The height profile measured along the line section in Figure 3a is shown in Figure 3d (black line). Decreasing the bias voltage to U = 100 mV we observed that the graphene was pushed against the top of the gold nanoisland (Figure 3b), as shown also by the corresponding height profile in Figure 3d (red line). Somewhat higher graphene z-values were obtained for the same height profile as the bias voltage was increased to U = 200 mV (Figure 3d, blue line). Furthermore, when the bias voltage was increased back to U = 1000 mV, the graphene bubble recovers to the initial shape (Figure 3d, green line), unlike the previous case (Figure 2). The current was kept constant throughout the measurements (I = 1 nA). These results show that by decreasing the bias voltage the repulsive force regime starts to dominate the tip-graphene interaction. The mechanical forces appearing below bias voltages of 200-250 mV are able to squeeze the graphene bubbles. Note that thermal buckling induced by tunnelling currents[23] can be excluded, since the current in our measurements was constant and relatively low (I = 1 nA).

In order to evaluate these forces between STM tip and graphene, we performed AFM measurements in PeakForce® mode[19] on similar graphene nanobubbles. The AFM image of a flat gold nanoisland with graphene bubble on top is shown in Figure 4a (*l*, *w* and *h* of 550 nm, 310 nm, and 43 nm). This AFM image was obtained by scanning with a very low (F = 1.5 nN) force. The height profile measured along the line section in Figure 4a (white line) shows that the height of the graphene bubble is about two times the height of the gold nanoisland (Figure 4d, black solid line). We performed AFM images of the same area using forces up to 40 nN. A complete image was recorded for every force setpoint (F). We observed that by increasing the scanning force the graphene bubble starts to collapse. This is shown in Figure 4b and Figure 4c, which are topographic images measured with forces of 20 nN and 35 nN, respectively. Obviously, the shape of the graphene bubble is altered.

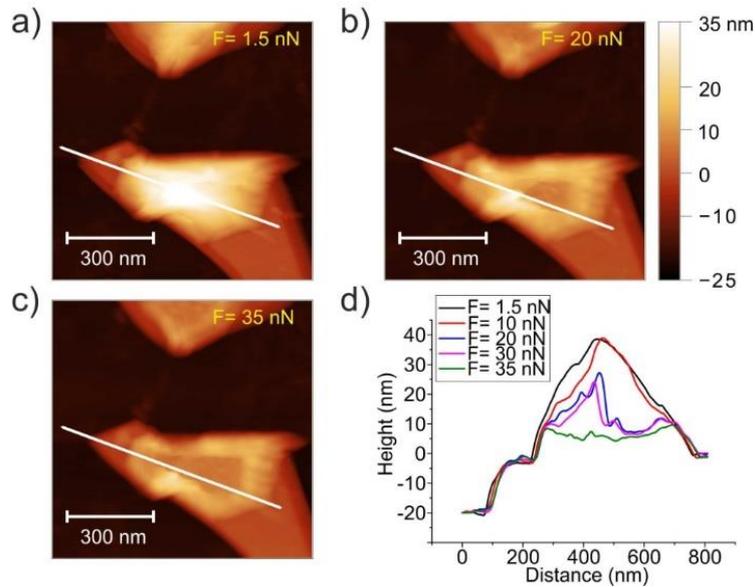

Figure 4. PeakForce AFM images of a graphene nanobubble measured with compressive forces of (a) F = 1.5 nN, (b) F = 20 nN, and (c) F = 35 nN. (d) Height profiles taken at different forces along the same line section shown in a)-c) (white line).

Selected height profiles are shown in Figure 4d, which were taken along the same line section shown in Figure 4a (white line), extracted from the AFM images measured with the corresponding tip-sample force values. The bubble is considered squeezed at F = 35 nN (Figure 4d, green line), when the shape becomes similar to the ones observed by STM at lower bias voltages (Figure 2c, Figure 3b).

*Graphene suspended on gold nanovoids*

Next, we investigated a sample which was prepared by evaporating 8 nm of gold on HOPG and transferring graphene on the evaporated gold.

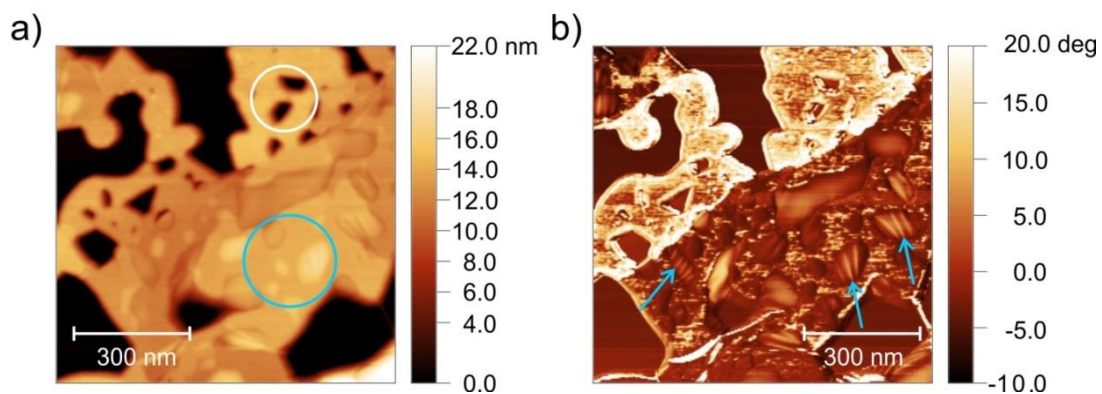

Figure 5. Tapping mode AFM images of gold nanostructures partially covered with graphene. (a) AFM topography. Two non-covered gold nanovoids are marked with white circle, while several graphene covered nanovoids are marked with blue circle. (b) AFM phase image of the same area. The light-coloured (upper image) parts correspond to bare gold nanostructures, while the darker contrast shows graphene covered gold nanostructures and nanovoids. The sharp boundary between light- and dark-coloured regions shows the edge of the graphene flake. The graphene covering the nanovoids is suspended and becomes rippled (blue arrows).

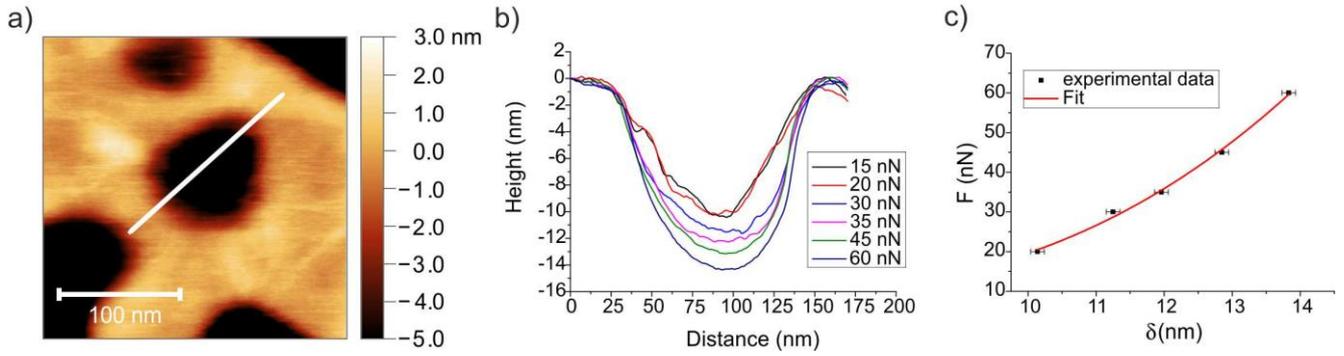

Figure 6. AFM nanoindentation performed in PeakForce mode. (a) Topographic image of a graphene covered gold nanovoid measured with a load force of F = 20 nN. (b) Height profiles taken at different forces along the same line section shown in a) (white line). (c) Force-deflection (δ) data obtained from b).

The sample was then annealed at 350 °C for 30 min in argon atmosphere. The formed gold nanostructures are flat and irregularly shaped, as shown by the AFM image in Figure 5a, measured in tapping mode after graphene transfer. In contrast to the previous sample, here the nanoislands are not well separated, but rather merged together to form larger structures. In this case, nanovoids of different shapes and sizes develop in the gold structure (see e.g. Figure 5a, white circle). Note that there are many gold nanovoids which are covered with graphene (Figure 5a, blue circle). During the AFM measurements we recorded also the phase shift signal from the same area, which is shown in Figure 5b. This phase image clearly shows the difference between uncovered and graphene-covered areas, which have higher- and lower phase contrast, respectively. The borderline between high- and low phase contrast (from top-right towards bottom-left of Figure 5b) gives the edge of the graphene flake covering the lower part of the image. The graphene parts which cover the nanovoids are suspended and several of these produce static rippling,[24] as revealed by the short parallel lines in the phase image (Figure 5b, blue arrows). Some of the graphene covered nanovoids appear as bright spots in the AFM image shown in Figure 5 (e.g. inside the blue circle). This convex buckling is probably due to some air trapped inside the voids during the graphene transfer, which can produce nanometer-sized blistering of graphene. Alternatively, the mismatch between the thermal expansion coefficients of graphene and gold can result in a compressive strain in graphene during annealing,[22] and hence a buckling.

We performed AFM nanoindentation experiments in PeakForce mode on graphene suspended over a gold nanovoid (not blistered), as shown in Figure 6. The same area (Figure 6a) was scanned at different peak forces and a complete image was recorded for every force setpoint (F). Figure 6b shows selected height profiles taken along the line section in Figure 6a (white line), extracted from the topographic images measured at the corresponding forces. The indentation-induced deflection (δ) was measured as the difference between minimum and maximum z-values of the height profiles. Force-deflection data were obtained, as shown in Figure 6c. Note that the suspended graphene has an initial deflection of $\delta_0$ = 10.1 nm, which starts to increase when F > $F_0$, where $F_0$ = 20 nN is a threshold force. For F < $F_0$, $\delta_0$ does not increase significantly.

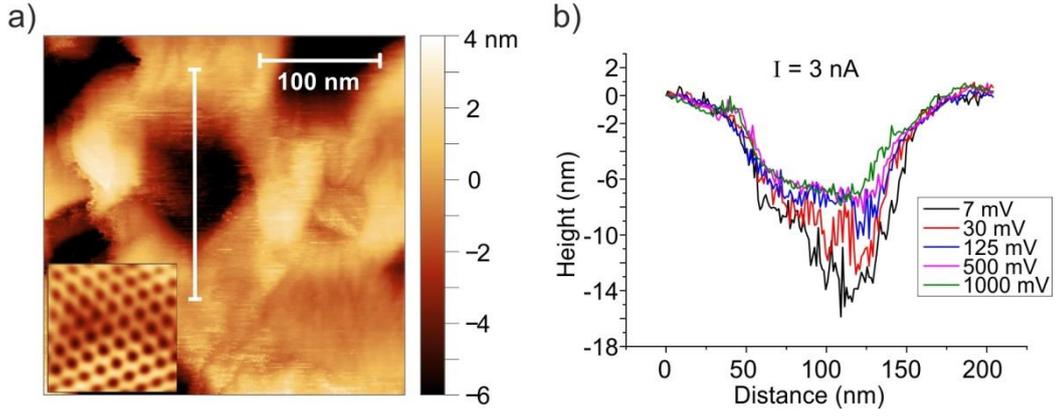

Figure 7. (a) STM image of a graphene covered gold nanovoid (U = 500 mV, I = 3 nA). The inset shows an atomic resolution STM image of graphene, measured on top of the gold nanostructures. (b) Height profiles taken at different bias voltages along the same line section shown in a) (white line).

To interpret the results we use the indentation model of a circular monolayer graphene using a rigid spherical indenter.[25] The suspended graphene area considered in Figure 6a can be approximated as a circular membrane with a radius of approximately a = 74 nm, while the radius of the AFM tip is R ≅ 12 nm. We fitted the force-deflection data with

$$F = c\delta + d\delta^3 \qquad (1)$$

where the coefficients $c$ and $d$ are related to the pre-tension and the Young's modulus (E) of a membrane of thickness h (0.34 nm for graphene), respectively.[25–28] In the case of the sphere load model ($R/a > 0.14$) $d = Eq^3 a^{-2} h (R/a)^{1/4}$,[29,30] where $q = 1/(1.05 - 0.15\nu - 0.16\nu^2) = 0.98$, and $\nu = 0.165$ is the Poisson's ratio for monolayer graphene.[26,31] From the fit (Figure 6c) we obtained E=0.95 TPa for the tensile modulus, which is in good agreement with previous findings on both CVD-grown[32,33] and exfoliated graphene.[26,34,35] Here, E was determined with a precision of 20 %, which comes from the uncertainty of the AFM tip radius, as well as from the deviation of the actual geometry of the suspended graphene from the circular shape. Note that the maximal deflection induced by indentation is $\delta_{max} - \delta_0 = 4.2\ nm$, obtained with a force of $\Delta F = F_{max} - F_0 = 40\ nN$, where $F_{max} = 60\ nN$ was the highest force used for AFM imaging.

Deflection of suspended graphene was observed by STM measurements as well. Figure 7a shows the STM image of a graphene covered gold nanovoid with dimensions similar to the one shown in Figure 6a. STM topographic images of the same area were measured in constant current mode (I = 3 nA) at different bias voltages. The height profiles along the white line in Figure 7a corresponding to different bias voltages are shown in Figure 7b. As we decrease the bias voltage below 500 mV repulsive forces appear and the deflection of the suspended graphene increases. A total deflection of $\delta \cong 7.5$ nm (black line in Figure 7b) is observed at U = 7 mV. Here, an initial deflection of $\delta_0$ = 7 nm observed at higher bias voltages (U > 500 mV) was already subtracted. The deflections measured by STM suggest that the repulsive forces associated with low bias voltages are comparable to the forces applied for AFM nanoindentation (Figure 6). We can evaluate the maximal force corresponding to the total deflection using the indentation model with a spherical indenter as above. We give an upper limit of r = 25 nm for the STM tip apex radius, estimated from the height profile measured at U = 7 mV (Figure 7b), which shows the deepest descend of the STM tip inside the gold nanovoid. Thus, r/a ≅ 0.31 for the graphene membrane considered in Figure 7 (a ≅ 80 nm), and the sphere load model[29,30] applies. Substituting δ = 7.5 nm in Equation 1 and neglecting the pre-tension, we obtain a force of F = 15 nN. Here, F was determined with a precision of 25 %, taking into account that the radius of the suspended graphene in Figure 7 deviates from the ideal circular geometry by around 10%. Nevertheless, the calculated STM tip-graphene force is in the same range as the forces used in the AFM experiments. In the calculation we used the tensile modulus E = 0.95 TPa determined above from AFM nanoindentation.

## Conclusions

Graphene covered gold nanovoids and graphene nanobubbles were investigated by STM in the repulsive force regime. We showed that the graphene nanobubbles could be squeezed during STM measurements using bias voltages of less than 250 mV and tunnelling currents of 1 nA. Similarly, the graphene suspended over gold nanovoids was deflected by the STM tip when imaging at low bias voltages. Nanoindentation measurements performed by AFM showed that the squeezing of graphene bubbles occurs at load forces of 20 – 35 nN, and such forces induced deflections of several nanometres in suspended graphene. Comparing the AFM and STM results we showed that repulsive forces of the order of $10^{-8}$ N occur between the STM tip and graphene under ambient imaging conditions and typical tunnelling parameters. More generally, the presence of such compressive forces can affect the STM measurements and should be taken into consideration during all low bias voltage investigations. Furthermore, our results allow the dynamic control of the local morphology of suspended graphene nanomembranes by both STM and AFM, which can open new pathways for the fabrication of nanomechanical devices.

## Acknowledgements

The research leading to these results has received funding from the People Programme (Marie Curie Actions) of the European Union's Seventh Framework Programme under REA grant agreement n° 334377. Support from the OTKA grant K-101599, the NKFIH project TÉT_12_SK-1-2013-0018, and from the Korea-Hungary Joint Laboratory for Nanosciences is acknowledged. Z. Osváth acknowledges the János Bolyai Research Fellowship from the Hungarian Academy of Sciences. C. Hwang acknowledges funding from the Nano-Material Technology Development Program (2012M3A7B4049888) through the National Research Foundation of Korea (NRF) funded by the Ministry of Science, ICT and Future Planning.